\documentclass[aps,prx, reprint,longbibliography, nofootinbib,superscriptaddress]{revtex4-2}

\usepackage{subfigure}
\usepackage{graphicx,amsmath,amssymb,color,amsfonts,bm,physics}
\usepackage{mathtools,lipsum}
\usepackage[dvipsnames]{xcolor}
\usepackage[misc]{ifsym}
\usepackage{quantikz}

\usepackage[colorlinks=true]{hyperref}  
\hypersetup{
    bookmarks=true,         
    unicode=false,          
    pdftoolbar=true,        
    pdfmenubar=true,        
    pdffitwindow=false,     
    pdfstartview={FitH},    
    pdftitle={},    
    pdfauthor={},     
    pdfsubject={},   
    pdfcreator={},   
    pdfproducer={}, 
    pdfkeywords={} {} {}, 
    pdfnewwindow=true,      
    colorlinks=true,       
    linkcolor=blue, 
    citecolor=blue,        
    filecolor=magenta,      
    urlcolor=blue,           
        breaklinks=true
} 

\newcommand{\be}{\begin{equation}}
\newcommand{\ee}{\end{equation}}
\newcommand{\bea}{\begin{eqnarray}}
\newcommand{\eea}{\end{eqnarray}}

\begin{document}

\title{Randomization Times under Quantum Chaotic Hamiltonian Evolution}

\author{Souradeep Ghosh}
\affiliation{Department of Physics \& Astronomy, Texas A\&M University, College Station, TX 77843}

\author{Nicholas Hunter-Jones}
\affiliation{Department of Physics and Department of Computer Science, University of Texas at Austin, Austin, TX 78712}

\author{Joaquin F. Rodriguez-Nieva}
\affiliation{Department of Physics \& Astronomy, Texas A\&M University, College Station, TX 77843}

\begin{abstract}

Randomness generation through quantum-chaotic evolution underpins foundational questions in statistical mechanics and applications across quantum information science, including benchmarking, tomography, metrology, and demonstrations of quantum computational advantage. 
While statistical mechanics successfully captures the temporal averages of local observables, understanding randomness at the level of higher statistical moments remains a daunting challenge, with analytic progress largely confined to random quantum circuit models or fine-tuned systems exhibiting `space–time' duality. Here we study how much randomness can be dynamically generated by {\it generic} quantum-chaotic evolution under {\it physical, non-random} Hamiltonians. Combining theoretical insights with numerical simulations, we show that for broad classes of initially unentangled states, the dynamics become effectively Haar-random well before the system can ergodically explore the physically accessible Hilbert space. Both local and highly nonlocal observables, including entanglement measures, equilibrate to their Haar expectation values and fluctuations on polynomial timescales with remarkably high numerical precision, and with the fastest randomization occurring in regions of parameter space previously identified as {\it maximally chaotic}. Interestingly, this effective randomization can occur on timescales {\it linear} in system size, suggesting that the sub-ballistic growth of Renyi entropies typically observed in systems with conservation laws can be bypassed in non-random Hamiltonians with an appropriate choice of initial conditions.

\end{abstract}



\maketitle


\noindent{\bf Introduction.---}Initially unentangled states evolving under quantum chaotic dynamics are generally expected to evolve into featureless states at late times. This expectation is supported by decades of theoretical and experimental research on quantum thermalization \cite{2011RMP_Quantumthermalization_review,2016AnnPhys_ETHreview,2019RMP_mblreview,baldovin2025foundations}, which has firmly established that the long-time, time-averaged behavior of local observables is accurately captured by thermal ensembles---even when the quantum system is not coupled to a thermal bath. In recent years, attention has increasingly shifted toward the structure of quantum states beyond local observables and beyond averages, focusing instead on universal features encoded in the fine-grained structure of quantum states. These questions are largely motivated by recent experimental advances in programmable quantum systems \cite{2019Nature_quantumsuppremacy,2021Nature_Rydberg256,2021RMP_trappedions,Periwal2021}, which enable direct access to fine-grained statistical information about quantum states—ranging from full counting statistics of measurement outcomes \cite{2023Nature_emergentdesign,Wienand2024} to nonlocal entanglement observables measured via novel measurement protocols \cite{2016Science_greiner,2020NatPhys_shadowtomography,2021PRL_randomquantumfisher,2023NatRev_randommeasurement,Joshi2023}.

In this spirit, a central question concerns the extent to which states generated by generic quantum-chaotic dynamics approximate Haar-random states, not merely in terms of ensemble-averaged observables, 
but also at the level of higher-order statistical moments. 
This question is of fundamental theoretical interest, as it underpins the notion of ergodicity---one of the central pillars of statistical mechanics---and has inspired the development of new ideas such as deep thermalization \cite{2023PRX_cotleremergentdesign,2023NatRev_randommeasurement} and complete Hilbert-space ergodicity \cite{2023PRL_completeergodicity}, together with statistical frameworks that characterize both the emergence of \cite{2018PRX_dualunitary,2019PRX_dualunitary}, and higher-order corrections to \cite{2019PRE_kurchan,Friedman2019,2024PRX_quantifyingchaos,PhysRevX.14.031029,PhysRevX.15.011031,2025PRL_EEU1,langlett2025quantumchaosfinitetemperature}, random matrix theory behavior in many-body quantum dynamics.
At the same time, Haar-random states constitute a powerful resource for quantum information science, as many quantum protocols and algorithms rely on access to highly entangled states, including tomography \cite{2020SIAM_shadowtomography,2020NatPhys_shadowtomography}, benchmarking \cite{Dankert09,2023PRL_chaosbenchmarking}, information recovery \cite{HaydenPreskillProtocol}, and demonstrations of quantum computational advantage \cite{2019Nature_quantumsuppremacy,2021Science_googlescrambling}. In practice, however, there are two major physical obstacles to realizing Haar randomness from local unitary evolution. 
First, ergodic exploration of Hilbert space requires exponentially long times \cite{PhysRevB.104.085117,PhysRevE.111.044210}, far exceeding the realistically accessible timescales of quantum systems. Second, conservation laws restrict the physically accessible Hilbert space: under Hamiltonian evolution, $|\Psi(t)\rangle = \sum_n e^{-i E_n t}|\langle n|\Psi_0\rangle||n\rangle$, the magnitudes of the 
projection of the initial state $|\Psi_0\rangle$ with the energy eigenstates are conserved, thereby preventing uniform exploration of the entire Hilbert space \cite{2024PRX_maximumentropyprinciple}.

In this work, we address how much randomness a local, non-random Hamiltonian can dynamically generate, and on what timescales. Our approach builds on two key recent insights. First, studies of random quantum circuit (RQC) models---where spatial and temporal randomness enable analytically tractable calculations---have shown that while achieving full Haar randomness requires exponential circuit depth, finite $k$-moments can be generated in linear-in-$L$ depth \cite{brandao2012local,chen2024incompressibility}. In other words, while generating truly Haar-random states requires a prohibitive number of gates, it is possible to reproduce Haar statistics at finite moments using a polynomial 
number of gates, forming an approximate unitary $k$-design and thereby retaining many of the useful properties of Haar-random states. 
Second, in Ref.~\cite{2025PRB_latetimeensemble} we showed that even when dynamics is constrained by conservation laws, which preclude uniform exploration of the full Hilbert space, time-evolved states can still generate the same finite $k$-moments of Haar random states provided the initial state has the same probability distribution of the conserved charge as the Haar ensemble. 

By combining these two insights with extensive numerical simulations, we find that: (i) non-random quantum-chaotic Hamiltonians can generate high degrees of randomness, such that all subsystem observables studied---from local observables to highly nonlocal entanglement observables---reproduce, to numerical precision, the corresponding moments of Haar-random states; (ii) these observables relax to their Haar expectation values and fluctuations on polynomial timescales, closely mirroring the behavior observed in RQCs, with the fastest randomization occurring in regions of parameter space identified as {\it maximally chaotic}; and (iii) for appropriate choices of initial conditions, the randomization time can scale as fast as linearly in system size $L$.

\begin{figure}
    \centering
    \includegraphics[scale=1.0]{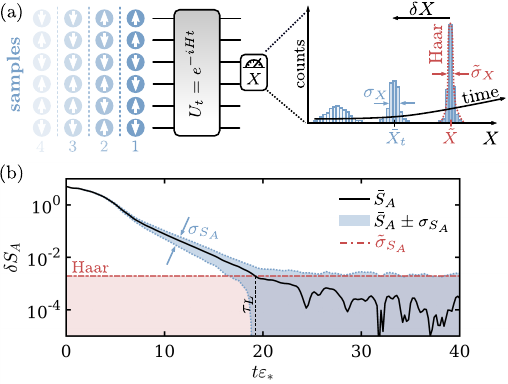}
    \vspace{-8mm}
    \caption{(a) Schematic of the theoretical setup. We sample initially unentangled states and evolve them under a generic quantum-chaotic Hamiltonian $H$. At each time $t$, we measure the full distribution of a chosen subsystem observable $X$ (e.g., $X:$ Renyi entropy, entanglement entropy, or subsystem energy). (b) Example: distribution of entanglement entropy $S_A$ of time-evolved states relative to the average entropy of Haar-random states $\tilde S_A$ ($\delta S_A = S_A - \tilde S_A$). Shown are the mean $\bar S_A$ over $M=100$ initial states (solid lines) along with the associated standard deviation $\sigma_{S_A}$ across samples (shaded region surrounding $\bar S_A$). 
    For comparison, we indicate with a shaded region the size of Haar fluctuations $\tilde{\sigma}_{S_A}$. 
    At times larger than $ \tau_L \approx 20/\varepsilon_*$, the evolved states reproduce both the mean and statistical fluctuations characteristic of Haar-random states ($\varepsilon_*$ denotes the characteristic local energy scale of $H$).}   
    \label{fig:schematics}
\end{figure}

\noindent{\bf Theoretical framework.}---The working definition of the randomization time we adopt is the time scale over which an initially unentangled state, evolving under a quantum-chaotic Hamiltonian, becomes effectively indistinguishable from a Haar-random state, specifically to an observer capable of measuring arbitrary subsystem observables up to some finite $k$-th moment. Importantly, our notion of indistinguishability is different from computational notions of indistinguishability \cite{schuster2025random}, where the observer is restricted to estimate observables using only a polynomial number of state copies (which excludes entanglement-based observables, and indistinguishability in that setting can arise on parametrically shorter, $O ( \log L)$, timescales \cite{schuster2025random,laracuente2024approximate}). 
With our definition, we aim to identify an {\it intrinsic} randomization time generated by quantum dynamics, independently of the observable of choice or the difficulty of measuring it. 

With this goal in mind, we define the {\it temporal ensemble} at time $t$ as
\be
\Phi_t = \big\{|\Psi_m\rangle: |\Psi_m\rangle = e^{-iHt}|\sigma_m\rangle , \,\,\, |\sigma_m\rangle \in \Phi_0\big\},
\label{eq:temporalensemble}
\ee
where the initial states $|\sigma_m\rangle$ are drawn from an ensemble $\Phi_0$ of unentangled states chosen to lie in the middle of the spectrum and having a `typical' energy variance (both of which will be specified more precisely below). The ensemble $\Phi_t$ consists of all such states evolved to a fixed time $t$. Starting from initially unentangled states is essential. Although the unitary evolution $e^{-iHt}$ simply rotates the states $|\sigma_m\rangle$ in Hilbert space without mixing them, entanglement diagnostics evaluated across samples provide an exceptionally sensitive probe of randomization, since the probability of returning to an unentangled state after generic quantum-chaotic evolution is exponentially small. 
As illustrated in Fig.~\ref{fig:schematics}(a), we probe $\Phi_t$ by sampling a variety of observables $X$ at different evolution times $t$, thereby generating a distribution $X_t$ of measurement outcomes.

A central aspect of our approach—departing from much of the prior work on quantum thermalization—is that we characterize randomization by directly comparing the distribution of $X_t$ against the {\it exact} distribution of $X$ for Haar-random states, as opposed to characterizing randomization solely through convergence to the late-time distribution $X_{t\rightarrow\infty}$. This distinction is crucial: we show that states do not merely {\it equilibrate} to their asymptotic values (captured by thermal ensembles), but in fact {\it randomize} in the stronger sense of reproducing the full Haar-level fluctuations of the observable $X$ (Fig.~\ref{fig:schematics}b).

Our approach is guided by our recent results showing that, although quantum states constrained by conservation laws do not explore the full Hilbert space, they can nevertheless become indistinguishable from Haar-random states at the level of finite 
statistical moments \cite{2025PRB_latetimeensemble}. Specifically, states generated by infinite-time evolution from an initial state $|\Psi_0\rangle$ give rise to the random-phase ensemble, 
$\Phi_{\rm RP}=\{ |\Psi_m\rangle=\sum_n e^{-i\theta_{mn}}|\langle n|\Psi_0\rangle||n\rangle,\theta_{mn}\in[0,2\pi)\}$, 
provided the Hamiltonian satisfies a no-$k$-resonance condition to all orders~\cite{2024PRX_maximumentropyprinciple}. 
While one might expect this ensemble to depend sensitively on the choice of $|\Psi_0\rangle$, we showed in Ref.~\cite{2025PRB_latetimeensemble} that $\Phi_{\rm RP}$ becomes, for all practical purposes, equivalent to the Haar ensemble at the level of finite statistical moments if the initial state matches Haar-random states in the moments of the conserved operator---i.e., $\langle\Psi_0|H|\Psi_0\rangle=\frac{1}{D}{\rm Tr}[H]$, $\langle\Psi_0|H^2|\Psi_0\rangle=\frac{1}{D}{\rm Tr}[H^2]$, and so on ($D$ is the Hilbert-space dimension). 
We established this equivalence exactly in U(1)-conserving random circuit models and verified it numerically in the largest accessible Hamiltonian systems. As illustrated in Fig.~\ref{fig:schematics}(b), for a moderate system size $L=16$ and $M=100$ samples, the distribution of the half-system entanglement entropy agrees with that of Haar random states, matching the mean value to five-digit precision and exhibiting identical, exponentially small state-to-state fluctuations.

\noindent{\bf Quantum chaotic evolution.}---We consider dynamics governed by the one-dimensional Mixed Field Ising Model (MFIM):
\begin{equation}
    {H} = \sum_{j} J {Z}_j{Z}_{j+1} + g {X}_j +h {Z}_j ,
    \label{eq:Ham}
\end{equation}
which has been widely studied as a canonical model of quantum chaotic dynamics~\cite{2011PRL_banulscirachastings, PhysRevLett.111.127205}. Here $X_j, Y_j, Z_j$ denote the Pauli matrices, $J$ is the exchange coupling, $g$ is the transverse field, and $h$ is the longitudinal field. Unless stated otherwise, we use $J=1$, $h=0.4$, and $g=1.05$. In addition, we use periodic boundary conditions, and break both translational and inversion symmetry with the choice of initial conditions, as discussed next.

Given $H$ in Eq.~(\ref{eq:Ham}), a natural choice of $\Phi_0$  
are product states oriented along the $\hat y$-axis, and arbitrarily pointing in the $\pm \hat y$ direction. Each of the states in $\Phi_0$ has: (i) zero energy, as $H$ does not contain any $Y_i$ operator, and (ii) an energy variance $\langle \sigma_m | H^2 | \sigma_m\rangle = \frac{1}{D}{\rm Tr}[H^2]$ matching exactly that of Haar random states. In addition, the ensemble defined this way forms a complete basis of the entire Hilbert space, providing us with an exponentially large sampling space to probe $\Phi_t$. 

In our plots we express energy and inverse time in units of $\varepsilon_*$, where $\varepsilon_*$ is defined from the variance of the Hamiltonian, ${\rm Tr}[H^2] = L\varepsilon_*^2$. Specifically, $\varepsilon_* = \sqrt{J^2+g^2+h^2}$ for $H$ in Eq.~\eqref{eq:Ham}. This allows us to compare randomization times across different model parameters, as this normalizes the typical level spacing.

\begin{figure}
    \centering
    \includegraphics[scale=1.0]{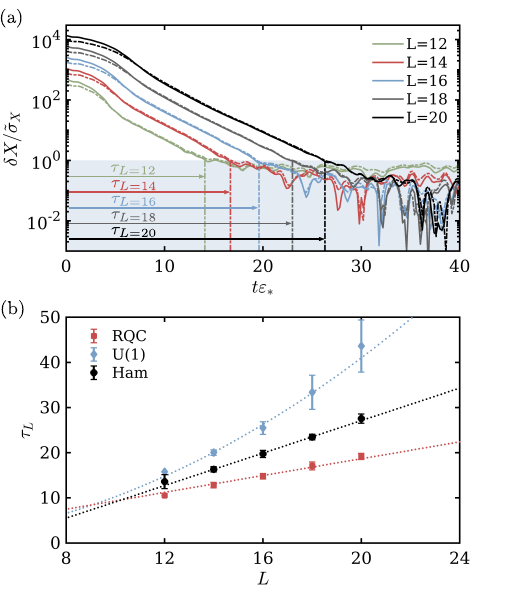}
    \vspace{-8mm}
    \caption{Quantum state randomization as measured through entanglement observables. (a) Relaxation of half-system entanglement observables to their Haar expectation values, shown for system sizes $L = 12, 14, \ldots, 20$ in the mixed-field Ising model (MFIM). We plot the deviation $\delta X(t) = \bar X(t) - \tilde X$, rescaled by the typical fluctuations $\tilde\sigma_X$. Data is shown for the entanglement entropy ($X=S_A$, solid lines) and second Renyi entropy ($X=S_2$, dash-dotted lines), with $\bar X$ averaged over 100 realizations, and with $\tilde X$ and $\tilde\sigma_X$ corresponding to the average and fluctuations of $X$ for Haar-random states. The shaded region serves as a guide to the eye to indicate the timescale where $\delta X = \tilde\sigma_X$. Parameters: $g = 1.05$ and $h = 0.4$. 
    (b) Randomization time $\tau_L$, defined as the timescale at which $\delta X(\tau_L) = \sigma_X$, obtained from $X=S_A$ and $X=S_2$ for subsystem sizes $L_A = 2,4,\ldots,L/2$ (contiguous sites). The bars indicate the variability of $\tau_L$ for across $X$ and $L_A$ values. For comparison, we show data of $\tau_L$ 
    in random quantum circuits (RQCs): $\tau_L \sim O(L)$ for circuits without a conservation law and $\tau_L \sim O(L^2)$ for U(1)-conserving RQCs.
}
    \label{fig:finitesizescaling}
\end{figure}

\noindent{\bf Randomization through the lens of entanglement.---}We first analyze randomization through entanglement observables, focusing on their {\it full distributions} 
including $O(1)$ corrections and fluctuations \cite{2022PRX_eereview}. 
Importantly, entanglement observables are extremely sensitive probes of randomization \cite{2020PRB_Grover_tripartite}: being nonlinear functions of the reduced density matrix $\rho_A$ of a subsystem $A$, they can detect extremely subtle, exponentially-small differences between two distinct quantum state ensembles---although this same nonlinearity makes them experimentally challenging to measure.

Returning to Fig.~\ref{fig:schematics}(b), the half-system entanglement entropy $S_A = -\mathrm{Tr}\left[\rho_A \log \rho_A\right]$,
exhibits two clearly distinct dynamical stages, as discussed in Ref.~\cite{2019PRL_EEevolution}. First, between microscopic $O(1)$ timescales and times of order $O(L_A)$, the entanglement entropy grows linearly in time until it approaches its volume-law value \cite{PhysRevLett.111.127205}. This is followed by a slower relaxation stage, during which the subleading structure of the entanglement spectrum---including its fluctuations and $O(1)$ corrections relative to the max-entropy---equilibrates to its stationary distribution \cite{2022PRA_Ranard}. We define the duration of this second stage as the randomization time $\tau_L$, and find that it scales polynomially with the total system size $L$, while remaining essentially independent of the subsystem size $L_A$ (see discussion below). Beyond this timescale, the agreement with Haar-random states is striking: Individual realizations exhibit temporal fluctuations that agree with remarkable precision with those of Haar-random states, both at the level of the first and second moments (indicated by solid and dotted lines, respectively). We find that this agreement holds for all Renyi entropies $S_n = \frac{1}{1-n}\log \mathrm{Tr}\left[\rho_A^n\right]$, including the min-entropy $S_\infty$. 

To highlight key features not visible in Fig.~\ref{fig:schematics}(b), we extend our analysis across multiple entanglement diagnostics, system sizes $L$, and subsystem sizes $L_A$ in Fig.~\ref{fig:finitesizescaling}. In Fig.~\ref{fig:finitesizescaling}(a), we plot the half-system von Neumann entropy $S_A$ and second Renyi entropy $S_2$ as functions of time for $L$ ranging from $L=12$ to $L=20$. A striking---and somewhat unexpected---feature is that both quantities equilibrate in tandem and in timescales that scale {\it linearly} with $L$, thereby suggesting that randomization occurs once the entanglement front propagates across the full system. This behavior persists for smaller subsystem sizes, as illustrated in Fig.~\ref{fig:finitesizescaling}(b), which shows little variability of $\tau_L$ (indicated by error bars) evaluated for both entanglement entropies and for different subsystem sizes $L_A \in [2,4,\ldots,L/2]$. Interestingly, although smaller subsystems reach their volume-law values earlier than larger subsystem, their fluctuations still require an $O(L)$ timescale to equilibrate to their exponentially small values. For comparison, we also include numerical results from 
RQC models \cite{2023AnnRev_randomcircuits}: circuits without conservation laws known to exhibit $\tau_L \sim O(L)$ 
randomization time \cite{HunterJones2019}, and circuits with a U(1) conservation law displaying `sub-ballistic' $\tau_L \sim O(L^2)$ scaling \cite{2018PRX_khemani,2019PRL_Rakovszky,Huang_2020}. In both cases, we employ brickwork circuits initialized in product states in the $x$-$y$ plane, which are the analog of the $y$-polarized initial conditions used in the MFIM, and rescale time such that all systems exhibit the same entanglement entropy ramp at $t < O(L_A)$.

The observed linear-in-$L$ randomization time is surprising, as it has been found that systems with conservation laws typically exhibit {\it sub-ballistic} growth of $S_n$ for $n \ge 2$, constrained by energy diffusion, which would instead suggest $\tau_L \sim O(L^2)$ \cite{Huang_2020,2019PRL_Rakovszky}. Analytical results supporting this expectation exist for U(1)-preserving quantum circuit models \cite{Huang_2020} and have been argued to extend to generic Hamiltonian dynamics \cite{2019PRL_Rakovszky}. However, a direct comparison between non-random Hamiltonian evolution and U(1)-conserving random circuits, shown in Fig.~\ref{fig:finitesizescaling}(b)—with both models exhibiting diffusion constants $D \sim O(1)$ \cite{2018PRX_khemani,2019PRX_lanczos}---reveals qualitatively distinct behavior at comparable system sizes. 
Similar scaling consistent with our findings were recently reported in a large-scale numerical study \cite{maceira2025thermalizationdynamicsclosedquantum}, focusing on the equilibration of local observables to their thermal values.

Although we cannot rigorously rule out the possibility that $S_A$ and $S_2$ in Fig.~\ref{fig:finitesizescaling}(a) ultimately separate into distinct asymptotic timescales---ballistic and diffusive, respectively---in the thermodynamic limit, we emphasize two key points. First, the initial states considered here are uniform in energy, and their local energy fluctuations are already `equilibrated' to their thermal values at $t=0$. Intuitively, there is no conserved charge that needs to be transported before entanglement can grow to its maximal value. This contrasts to the setup of Ref.~\cite{2019PRL_Rakovszky}, where random product-state initial conditions generate spatially inhomogeneous energy fluctuations that must first relax diffusively before equilibration can occur. 

Second, a commonly used argument in the U(1)-conserving RQCs is that even when the initial state has a uniform charge density, it nonetheless overlaps with low-density, `domain-wall' states that slowly equilibrate diffusively and therefore bound the decay rate of the largest eigenvalue of the entanglement spectrum~\cite{Huang_2020}. While this reasoning is appropriate for RQCs---whose low-density charge sectors are diffusive by construction---it may not apply to non-random Hamiltonians. In particular, low-energy sectors of generic one-dimensional Hamiltonians are typically ballistic, as described by Luttinger-liquid theory, and therefore need not impose a diffusive bottleneck on entanglement growth.

\begin{figure}
    \centering
    \includegraphics[scale=1.0]{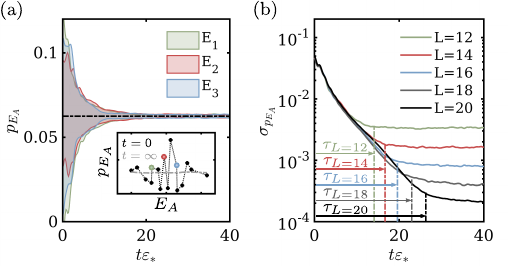}
    \vspace{-6mm}
    \caption{Dynamics of the full counting statistics of energy within a subsystem of size $L_A = 4$ embedded in a chain of $L = 16$ spins. 
    (a) Shown is the distribution of $p_{E_A}$ for three representative energy levels $E_A$ and $M=100$ initial conditions, illustrating exponential-in-time convergence to their Haar-like values (dotted-dashed lines) on the same timescale $\tau_L$ identified in Fig. \ref{fig:finitesizescaling}. The shaded regions indicate the variance $\sigma_{p_{E_A}}$ of $p_{E_A}$ across different initializations. The inset shows the distribution $p_{E_A}$ for the 16 energy levels in subsystem $A$ at time $t=0$ and in the long-time limit $t \to \infty$, for a single initialization. 
    (b) System-size scaling of the timescales at which the fluctuations of $p_{E_A}$ reach their late-time value, averaged over all energy outcomes $E_A$ in $A$.
    }
    \label{fig:fullcounting}
\end{figure}

\noindent{\bf Randomization time through full-counting statistics of energy.---}As an independent check of the $O(L)$ scaling of the randomization time, we examine the full counting statistics (FCS) of energy in a subsystem \cite{2023PRL_fullcountingstatistics,2024PRB_fullcountingstatistics,Wienand2024}. 
Specifically, we consider the local energy $H_A = \sum_{i=1}^{L_A}\big(h Z_i + g X_i\big) + \sum_{i=1}^{L_A-1} J Z_i Z_{i+1}$ for a subsystem $A$, and study the time evolution of the corresponding probability distribution $p_{E_A}$ over energy outcomes $E_A$.

In Fig.~\ref{fig:fullcounting}(a), we show the sample-to-sample fluctuations of $p_{E_A}$ for three representative energy outcomes and $L_A=4$, sampled over $M=100$ initial conditions. As discussed above, the states in $\Phi_0$ have mean $p_{E_A}$ already equilibrated to its late-time Haar average (dotted-dashed line), yet the distribution of $p_{E_A}$ fluctuates in time around this value, which is indicated by shaded regions.
We find that these temporal fluctuations become equal to those of  Haar-random scale on the same $O(L)$ timescales as those inferred from entanglement diagnostics, with no signatures of slow hydrodynamic relaxation. Panel (b) further supports this conclusion: averaging over all energies $E_A$,  we observe the same exponential approach to Haar randomness seen in the entanglement observable for all accessible system sizes. The corresponding randomization times---indicated by vertical dash-dotted lines---coincide with those extracted from Fig.~\ref{fig:finitesizescaling}, again showing no evidence of diffusive or otherwise parametrically slower modes. 

\begin{figure}
    \centering
    \includegraphics[scale=1.0]{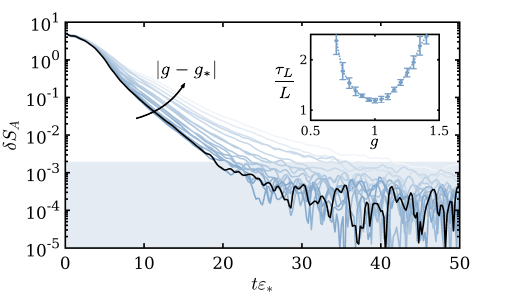}
    \vspace{-8mm}
    \caption{Randomization of time evolved states, as measured through the relaxation of the entanglement entropy distribution, 
    as a function of the transverse field $g$ in the MFIM.     Results are shown for $M = 10^2$ samples of the half-system entanglement entropy in a chain of size $L = 16$, with $g$ varied in increments of $\Delta g = 0.05$. The black curve corresponds to the parameter value with the fastest randomization, $g_* = 1.00\pm 0.05$, while the progressively lighter shades denote increasing $|g - g_*|$. The shaded region indicates the typical entanglement fluctuations expected for Haar-random states. The inset displays the randomization time as a function of $g$, normalized by system size. }
    \label{fig:gdependence}
\end{figure}

\noindent{\bf Randomization time across Hamiltonians.---}We now examine how sensitive $\tau_L$ is to variations in the Hamiltonian parameters. We find that $\tau_L$ depends strongly on the model parameters and is minimized in regions previously identified as {\it maximally chaotic}.

Specifically, Fig.~\ref{fig:gdependence} shows the equilibration of 
$S_A$, analogous to Fig.~\ref{fig:schematics}, but plotted for different values of the transverse field $g$.  
Sweeping $g$ over a broad range, $g\in[0,2]$, and extracting the corresponding randomization times $\tau_L$ using the same procedure as in Fig.~\ref{fig:finitesizescaling}, we find that $\tau_L$ varies smoothly with $g$ and exhibits a pronounced minimum at $g_* = 1.00 \pm 0.05$, as highlighted by the black curve in Fig.~\ref{fig:gdependence} (see inset). Remarkably, this optimal value of $g$ agrees, within uncertainties, with the parameter regime previously identified as maximally chaotic based on diagnostics that quantify {\it eigenstate} randomness \cite{2024PRX_quantifyingchaos}. Our results therefore provide an independent, fully dynamical signature of maximal chaos, obtained directly from the real-time evolution of initially unentangled product states.

\noindent{\bf Discussion.---}Taken together, our results reveal three independent and striking features of dynamics under non-random quantum-chaotic Hamiltonian evolution. First, for broad classes of unentangled initial conditions, we find that late-time states do not merely {\it thermalize} to their ensemble-averaged asymptotic values predicted by the Gibbs ensemble, but 
they randomize in a much stronger sense, reproducing Haar-level fluctuations of any subsystem observable $X$. Importantly, this holds not only for local observables and their temporal average, as typically considered in the quantum thermalization literature, but also for highly nonlocal 
quantities, including half-system entanglement entropies and their fluctuations. 

Second, we show that the randomization time is polynomial in the number of spins. In other words, it is not necessary to wait exponentially long  
times for the state to explore the entire accessible Hilbert space and exhibit Haar-like behavior but, instead, this can occur much earlier. 
This behavior resembles that of random quantum circuits, 
where exponential-in-time convergence of the $k$-moments to their Haar values has been established analytically. It is therefore striking that qualitatively similar behavior emerges from the dynamics of non-random, translationally invariant energy-conserving systems. 

Third, and perhaps most surprisingly, for the initial states considered in this work---those with uniform energy and `equilibrated' energy fluctuations, we find that unentangled states can effectively randomize in {\it linear-in-$L$} times for all the numericall-accessible system sizes. This behavior is consistently observed across all classes of local and nonlocal observables.  
This appears to contradict prior results suggesting that Renyi-entropy growth in RQCs with U(1) conservation is constrained by diffusive hydrodynamic modes.   
We provided plausible arguments that reconcile these differences, which we will explore further in upcoming work. As they stand, our results suggest that nonrandom Hamiltonians can effectively randomize faster than their charge-conserving RQCs counterparts with proper choice of initial states. 

\noindent{\bf Acknowledgments.---}We thank Wen Wei Ho, Anatoli Polkovnikov, Federica Surace, Romain Vasseur, Sagar Vijay, and Tianci Zhou for insightful discussions. JRN and NHJ acknowledges the hospitality of the Kavli Institute for Theoretical Physics through the program {\it Learning the Fine Structure of Quantum Dynamics in Programmable Quantum Matter}, supported by NSF Grant No.\ PHY-2309135. JRN also acknowledges the hospitality of the Aspen Center for Physics, which is supported by NSF Grant No.\ PHY-2210452 and by a grant from the Alfred P. Sloan Foundation (G-2024-22395). NHJ acknowledges support from DOE Grant DE-SC0025615. Numerical simulations were performed using the advanced computing resources provided by Texas A\&M High Performance Research Computing.

%

\end{document}